\documentclass[11ptUSenglish,onecolumn]{article}
\usepackage[a4paper, total={6in, 8in}]{geometry}

\usepackage[utf8]{inputenc}
\usepackage{natbib}
\usepackage{times}
 
\usepackage{hyperref}
\usepackage{xcolor}		
\definecolor{darkblue}{rgb}{0, 0, 0.5}
\hypersetup{colorlinks=true,citecolor=darkblue, linkcolor=darkblue, urlcolor=darkblue}
\usepackage{amsmath}
\usepackage{graphicx}
\usepackage{tikz}
\usepackage{amssymb}
\usetikzlibrary{arrows}
\usepackage{hyperref} 
\usepackage{amsthm}
\usepackage{float}
\usepackage{mathrsfs} 
\usepackage[final]{pdfpages}
\usepackage[normalem]{ulem}
\usepackage{setspace}
\usepackage{enumitem}   
\usepackage{algorithm,algorithmic}
\usepackage{caption}

\makeatletter

      \theoremstyle{plain}

\newcommand{\expit}{\text{expit}}

\newcommand\@erelb@r[1]{%
  \mathrel{\tikz[baseline=-.5ex]\draw[#1] (0,0)--(0.3,0);}
}

\newcommand{\erelbar}[1]{\@erelbar#1}
\def\@erelbar#1#2{%
  \ifcase\numexpr#1*4+#2\relax
    \@erelb@r{-}\or     
    \@erelb@r{->}\or    
    \@erelb@r{-|}\or    
    \@erelb@r{->|}\or   
    \@erelb@r{<-}\or    
    \@erelb@r{<->}\or   
    \@erelb@r{<-|}\or   
    \@erelb@r{<->}\or   
    \@erelb@r{|-}\or    
    \@erelb@r{|->}\or   
    \@erelb@r{|-|}\or   
    \@erelb@r{|<->|}\or 
    \@erelb@r{|<-}\or   
    \@erelb@r{|<->}\or  
    \@erelb@r{|<-|}\or  
    \@erelb@r{|<->|}    
  \else
    \@wrong
  \fi
}
\makeatother

\newcommand{\gol}{
\overset{\mathcal{L}}{\to}
}

\newlength\myindent
\newcommand{\AC}{\mathcal{A}}

\newcommand{{\citerelativesparsity}}{\cite{relsparSIM}}
\newcommand{{\citeinference}}{our work}
\usepackage{authblk}

  \title{\Large An adaptive variance estimator for relative sparsity}
\author{Samuel J. Weisenthal}
\affil{\text{samweisenthal@gmail.com}}

\begin{document}











\maketitle

 \begin{abstract}
An approach to inference for relative sparsity  was developed in prior work, and an adaptive lasso asymptotic normality theorem was given there, but this theorem was not fully used when estimating the  variance of the policy coefficients. Here, we develop a new coefficient variance estimator that fully uses this theorem and, in the process, takes into account the variable selection.  This improves the uncertainty representation in the graphical selection diagrams, ultimately facilitating the safe use of policy learning in clinical medicine.  
\end{abstract}


\section{Introduction}


The relative sparsity penalty \citep{relsparSIM}, building on Trust Policy Optimization \citep{schulman2015trust}, was developed as a way to derive a policy with a sparse difference from the standard of care, facilitating interpretation and justification of a shift from one policy to another (which can be particularly valuable for assessing policy changes that depend on assumptions about confounding in offline learning).  Following this, \citet{weisenthal2023inference} developed a framework for inference for relative sparsity.   A version of the adaptive lasso \citep{zou2006adaptive} theorem for the relative sparsity penalty was given in \citet{weisenthal2023inference}, but it was not fully used, and the selection diagram coefficient variance estimates were therefore poor in the behavioral region. Here, a new, adaptive coefficient variance estimator is proposed, which makes use of the adaptive relative sparsity theorem and, in doing so, takes into account the selection. Results for the proposed estimator are compared with \citet{weisenthal2023inference} with respect to the simulation studies and real data analysis (where the latter uses the MIMIC III database \citep{johnson2016mimic2, johnson2016mimic1, goldberger2000physiobank}). 

In general, data-driven treatment policies are an exciting research area. For example, they have potential to aid in clinical problems including the management of hypotension and sepsis \citep{relsparSIM, raghu2017continuous, saria2018individualized, kalimouttou2025optimal}, addiction \citep{jones2022valid}, and percutaneous endoscopic gastrostomy tube placement \citep{lu2017direct}. Constrained treatment policies---such as the ones derived via the relative sparsity penalty---might facilitate their safe adoption, which will further be made possible by the estimator derived here. 

\section{Background}
 \label{sec:not}
We use the same notation and setup as \citet{weisenthal2023inference},  in which a multi-stage, discrete-time, stationary Markov decision process \citep{bellman1957markovian} was considered (with state $S\in \mathbb{R}^K$ and action, $A\in\{0,1\}$, for $n$ independent and identically distributed length-($T+1$) trajectories). As in \citet{weisenthal2023inference}, a parameterized suggested policy, $\pi_{\beta},$ and behavioral policy, $\pi_b,$ which were expit functions in their respective parameters, will be used. We will also use the same deterministic, stationary reward function, $R(S_t,A_t,S_{t+1}),$ whose expectation is value, $V_0.$ 

As in \citet{weisenthal2023inference}, we will use the base objective $M_{0}(\beta,b,\gamma)=V_0(\beta) - \gamma KL_0(\beta,b),$ where $KL_0$ is KL-divergence \citep{kullback1951information}, which gives the adaptive relative sparsity penalty $W_0(\beta,\beta_{0,\gamma},b_0)=M_{0}(\beta,b,\gamma)-\lambda\sum_{k=1}^K{w}_{0,k} |\beta_k-b_{0,k}|$ with estimand $\beta_{0,\gamma,\lambda}=\arg\max_{\beta} W_0(\beta,\beta_{0,\gamma},b_0).$  Also as in \citet{weisenthal2023inference}, define $\AC$ to be a set containing the indices of selected covariates, $1_{\AC}$ to be its complement, $1_{\mathcal{A}}=(1_{1\in \mathcal{A}},\dots,1_{K \in \mathcal{A}})^T,$ and,  if $\odot$ denotes element-wise multiplication, define $\pi_{\beta,b}(1|s)=\expit(\beta^T (s\odot 1_{\mathcal{A}})+b^T (s\odot (1_K-1_{\mathcal{A}})) ),$ where $1_K$ is a length-$K$ vector of ones. As in \citet{weisenthal2023inference}, the parameter $b_0$ will be estimated with $b_n,$ $V_0$ with importance sampling \citep{precup2000eligibility, thomas2015safe} as ${V}_n(\beta,b_n),$ and $M_0(\beta,b,\gamma)$ with ${M}_n(\beta,b_n,\gamma)$ using $V_n(\beta)$ and  $KL_n(\beta,b_n).$  Further, as in \citet{weisenthal2023inference}, $\beta_{0,\gamma}$ will be estimated with $\beta_{n,\gamma}=\arg\max_{\beta}{M}_n(\beta,b_n,\gamma),$ $W_0$ with  $W_n(\beta,\beta_{n,\gamma},b_n)={M}_n(\beta,b_n)-\lambda\sum_{k=1}^K w_{n,k} |\beta_k-b_{n,k}|,$ where $w_{0,k}$ is estimated with $w_{n,k}={1}/{|\beta_{n,\gamma,k}-b_{n,k}|^{\delta}},$ and $\beta_{0,\gamma,\lambda}$ with $\beta_{n,\gamma,\lambda}=\arg\max_{\beta}W_n(\beta,\beta_{n,\gamma},b_n).$

As in \citet{weisenthal2023inference}, we make the necessary causal assumptions (positivity, consistency, no interference, and sequential randomization) and asymptotic assumptions (the parameters $\beta$ and $b$ exist in a compact set,  the states are bounded,  the reward is bounded,  the partial derivatives of $M_0$ are bounded by an integrable and measurable function,  $M_0$ has a unique maximizer, and  $b_n$ is consistent for $b_0$). Finally, as in \citet{weisenthal2023inference}, we define, with respect to $M_n,$ a gradient, $J_0,$ and its estimator, $J_n,$ a hessian, $H_0,$ (assuming it exists and is non-singular) and its estimator, $H_n,$ and a cross derivative, $X_0,$ and its estimator, $X_n.$  

Recall that \citet{weisenthal2023inference} showed consistency and asymptotic normality of $\beta_{n,\gamma}$ in Theorem 1 and, in Theorem 2, for $\delta>0,$ $\frac{\lambda_n}{\sqrt{n}}\rightarrow 0$ and $\lambda_n n^{(\delta-1)/2}\rightarrow \infty,$ active set, $\AC=\{k:{(\beta_{0,\gamma,k}}-b_{0,k})\neq 0\},$ and the indexed coefficients, $\beta_{0,\gamma,\AC},$ it was shown that, for $r_{v_0}=[z_{0,\AC}^T+u_{0,\AC^C}^TH_{0,\AC^C\AC}+u_{0,\AC^C}^T{(}H_{0,\AC\AC^C}{)^T}+v_{0,\AC}^T(X_0^{T})_{\AC\AC}+v_{0,\AC^C}^T(X_0^{T})_{\AC^C\AC}]^T,$  \[\sqrt{n}(\beta_{n,\gamma,\AC}-\beta_{0,\gamma,\AC})\gol N(0,(H_{0,\AC\AC})^{-1}\text{var}(r_{v_0}^T)((H_{0,\AC\AC})^{-1})^T).\] This result will be the basis for the proposed estimator here.

\section{Adaptive variance estimator}

We now propose a variance estimator for the adaptive Lasso coefficients, based on the aforementioned Theorem 2 of \citet{weisenthal2023inference}. Recall that $\AC$ and $\AC^C$ index the parameters that are selected/active/non-behavioral and unselected/non-active/behavioral, respectively, and that an empirical selection is represented as $\AC_n$ or $\AC_n^C$.   Theorem 2 gives us a form for the variance similar to the one given in Theorem 1 in \citet{weisenthal2023inference}, and we follow arguments that are analogous to those made in the derivation of (25) in Appendix A.17 of \citet{weisenthal2023inference} to propose an adaptive Lasso coefficient variance estimator,
\begin{equation}
\label{eq:sigma2nAdaptive}
\sigma^2_{n,A} = ((H_n)_{\AC_n\AC_n})^{-1} \left(\frac{1}{n}\sum_i (r_{v_{n,i}}^T)(r_{v_{n,i}}^T)^T\right) \left(((H_n)_{\AC_n\AC_n})^{-1}\right)^T.
\end{equation}
We estimate $r_{v_0},$ defined in Theorem 2 of \citet{weisenthal2023inference}, with $r_{v,n}=\frac{1}{n}\sum_ir_{v,n,i},$ where for observation $i$
\begin{align*}
r_{v_{n,i}} &=
[z_{n,i,\AC_n}^T+u_{n,i,\AC_n^C}^T(H_n)_{\AC_n^C\AC_n}+u_{n,i,\AC_n^C}^T((H_n)_{\AC_n\AC_n^C})^T+v_{n,i,\AC_n}^T(X_n^T)_{\AC_n\AC_n}+v_{n,i,\AC_n^C}^T(X_n^T)_{\AC_n^C\AC_n}]^T\\
&=[z_{n,i,\AC_n}^T+v_{n,i,\AC_n^C}^T(H_n)_{\AC_n^C\AC_n}+v_{n,i,\AC_n^C}^T((H_n)_{\AC_n\AC_n^C})^T+v_{n,i,\AC_n}^T(X_n^T)_{\AC_n\AC_n}+v_{n,i,\AC_n^C}^T(X_n^T)_{\AC_n^C\AC_n}].^T
\end{align*}
We were able to replace $u_{A^C}$ with $v_{A^C}$ in the last display, because the behavioral components of $u$ are equal to their behavioral counterparts in $v$. Recall that the nuisance is $v=\sqrt{n}(b_n-b_0),$ and we have an estimator $v_{n,i}=q_i,$ where $q$ is defined in (24) of \citet{weisenthal2023inference}. As in \citet{weisenthal2023inference}, this estimator (\ref{eq:sigma2nAdaptive}) can be used to visualize the variability of $\beta_{n,\gamma,\lambda}$ in the training data. 

The estimator for the variance of the value, $V_n,$ shown in \citet{weisenthal2023inference} (27),  remains unchanged (we display it in the selection diagrams here because it was displayed in \citet{weisenthal2023inference}).  We follow the estimation algorithm described in Section 7.3 of \citet{weisenthal2023inference}.

\section{Simulation}
 We use the simulation scenario, data generation per \citet{ertefaie2018constructing}, and reward function, $R(s_t,a_t,s_{t+1})=-s_{t,2}a_t,$ from Section 8 of \citet{weisenthal2023inference}. 
\label{sec:sim}

\begin{figure}
\centering
\includegraphics[width=0.98\textwidth]{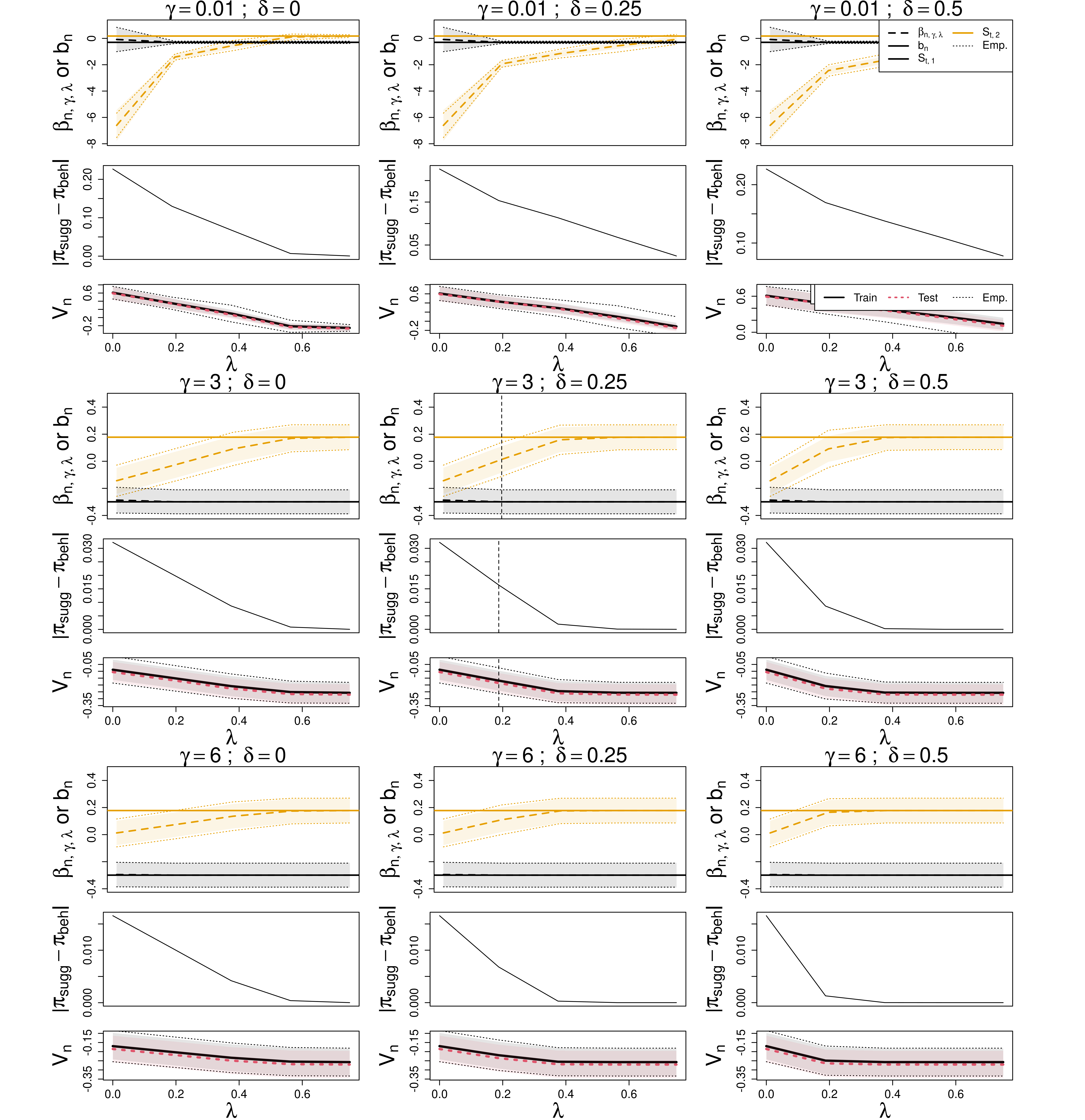}
	\caption[Selection diagrams for the simulated data]{
\textbf{Selection diagram.} We show an identical selection diagram to the one in Figure 1 of \citet{weisenthal2023inference}, but the estimator for the variance of the coefficients is now (\ref{eq:sigma2nAdaptive}) rather than equation (25) of \citet{weisenthal2023inference}. Recall that the shaded regions in the coefficient panels correspond to the theoretical variances (using (\ref{eq:sigma2nAdaptive}) here), and the dotted lines to the empirical variances (one standard error).
 }
\label{fig:sim}
\end{figure}

We see that the coefficient variance estimators in Figure \ref{fig:sim} show different behavior for small $\gamma$ (the top three panels) when compared to the corresponding selection diagram in Figure 1 of \citet{weisenthal2023inference}, specifically as they approach and then equal their behavioral counterparts, $b_n$  (i.e., as  $\lambda$ increases).  In particular, around this behavioral region, the estimated variance  is more controlled.

\section{Real data analysis}
\label{realdata}

\begin{figure}
\centering
\includegraphics[width=0.97\textwidth]{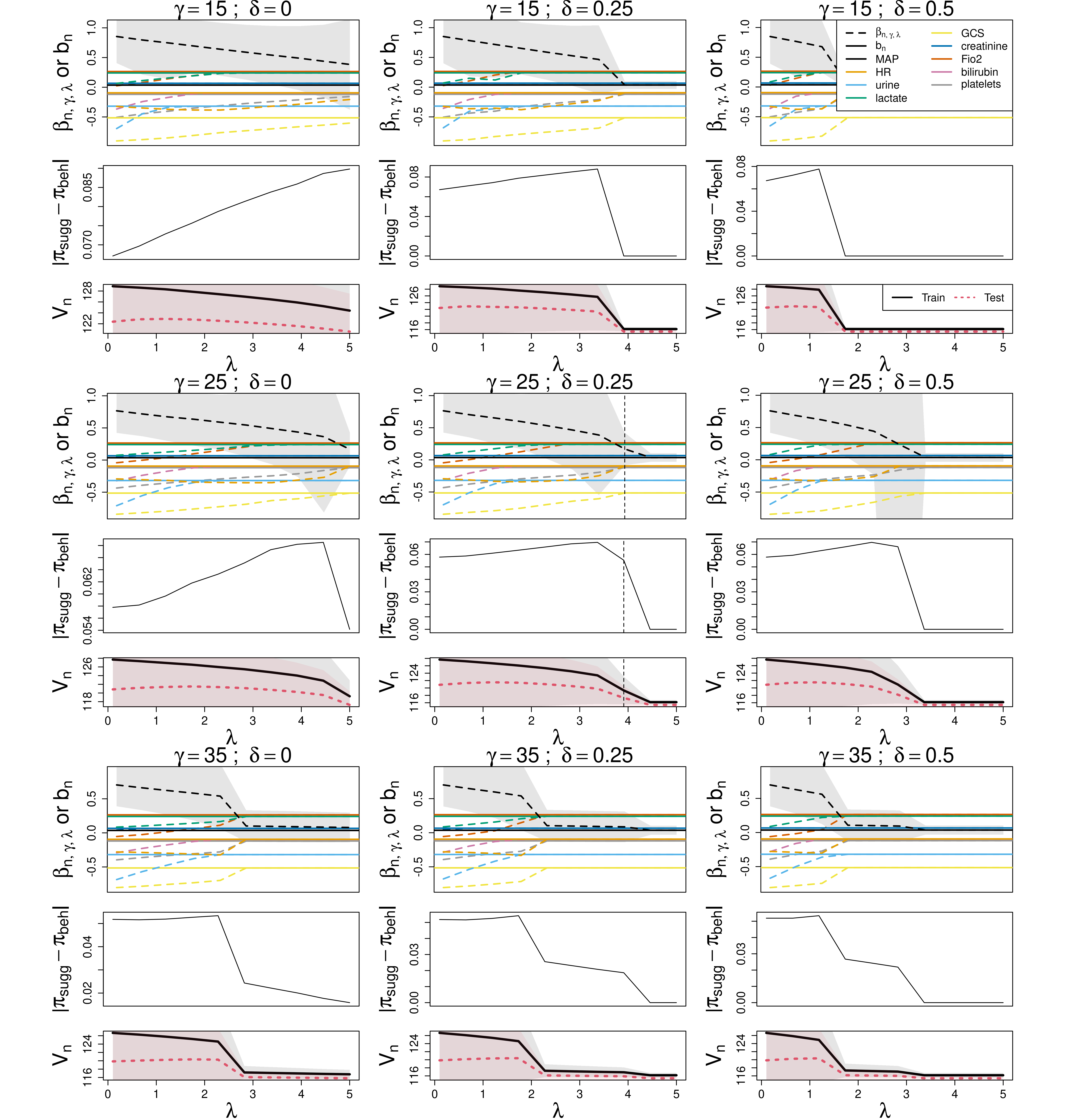}
\caption[Selection diagrams for the real data]{{\bf Selection diagrams for the real data.}   We show an identical real data selection diagram to the one in \citet{weisenthal2023inference}, but the estimator for the variance of the coefficients is now (\ref{eq:sigma2nAdaptive}) rather than (25) in \citet{weisenthal2023inference}. Recall that the shaded regions in the coefficient panels correspond to the theoretical variances (using (\ref{eq:sigma2nAdaptive}) here), and the dotted lines to the empirical variances (one standard error). As in \citet{weisenthal2023inference}, we show the variance only for mean arterial pressure, since it was selected.} 
\label{fig:real.data}
\end{figure}

We also show how the proposed  estimator, (\ref{eq:sigma2nAdaptive}), compares to the estimator (25) in \citet{weisenthal2023inference} in real data. We consider the same medical decision problem described in Section 9.1 of \citet{weisenthal2023inference}, which was a multi-stage version of the decision problem in \citet{relsparSIM}, and focused on vasopressor administration for hypotension in the medical intensive care unit, making use of existing pre-processing code and MIMIC data \cite{futoma2020popcorn, johnson2016mimic2, johnson2016mimic1, goldberger2000physiobank}.  In particular, recall that \citet{weisenthal2023inference} considered a decision problem with the goal of increasing mean arterial pressure within 45 minutes, where the initial state included the patient covariates from \citet{futoma2020popcorn}, which were mean arterial pressure (MAP), heart rate (HR), urine output (Urine), lactate, Glasgow coma scale (GCS), serum creatinine, fraction of inspired oxygen (FiO2), total bilirubin, and platelet count. Also, recall that the action corresponded to vasopressor administration, and the reward corresponded to the final mean arterial pressure. Recall that the behavioral policy estimator was found to be reasonably calibrated as shown in Figure A.1 of \citet{weisenthal2023inference}.

As in the simulations, we see in Figure \ref{fig:real.data} that, compared to Figure 2 of \citet{weisenthal2023inference}, the behavior is markedly different in almost all panels; in particular, as we saw in Section \ref{sec:sim}, the estimated variance is better controlled in the regions in which the estimated coefficients are nearer to their behavioral counterparts (i.e., as  $\lambda$ increases).

\section{Discussion}
\label{sec:discussion}
In this work, an adaptive relative sparsity coefficient variance estimator has been proposed, making use of results from \citet{zou2006adaptive} and \citet{weisenthal2023inference}.

In \citet{weisenthal2023inference}, although an approach for inference for relative sparsity was developed, only the variance estimator for the active coefficients was used in the selection diagram. This made partial use of Theorem 2 of \citet{weisenthal2023inference}---since Theorem 2 implied that the active coefficient variance is a reasonable estimator for small $\lambda$---but yielded poor estimates for coefficients near their  behavioral counterparts.  The proposed estimator makes fuller use of Theorem 2, and, in doing so, the selection itself. In particular, we see selection-specific coefficient variance estimates related to either the trust region policy optimization coefficients, $\beta_{n,\gamma},$ derived by maximizing $M_n$ alone, or behavioral coefficients, $b_n,$ obtained by maximizing $KL_n$ alone.  This leads to better controlled estimators for coefficients that are nearer to their behavioral counterparts, as shown here in  simulations (Figure \ref{fig:sim}) and real data (Figure \ref{fig:real.data}).

Overall, it is best to visualize the variability of the coefficients under changing $\lambda$---and the corresponding changing selections---perhaps on a fine grid. The proposed estimator facilitates this. In the future, one might consider  studying the properties of the proposed variance estimator as a smooth function of $\lambda.$ 

 Overall, such techniques, applied to different datasets, can facilitate the safe use of policy learning in clinical medicine.


\section{Acknowledgments}
This work extends the thesis \citep{weisenthal2023relative}, and the author acknowledges the faculty, post-doctoral, and graduate researchers who played a role in it. Further, after the thesis, S. W. Thurston provided helpful comments on some of what is written/shown here.


\bibliographystyle{plainnat}
\bibliography{referencesnew}


\appendix
 \renewcommand\thefigure{\thesection.\arabic{figure}}  
 \setcounter{figure}{0}   

\end{document}